# 6G V2X Technologies and Orchestrated Sensing for Autonomous Driving

## A vision on communication and sensing technologies, applications, and research trends for the future of intelligent mobility services

Marouan Mizmizi[1], Mattia Brambilla[2], Dario Tagliaferri[1], Christian Mazzucco[3], Merouane Debbah[4], Tomasz Mach[5], Rino Simeone[6], Silvio Mandelli[7], Valerio Frascolla[8], Renato Lombardi[3], Maurizio Magarini[1], Monica Nicoli[2], Umberto Spagnolini[1]

[1] Dipartimento di Elettronica, Informazione e Bioingegneria, Politecnico di Milano, 20133 Milan, Italy (name.surname@polimi.it)
[2] Dipartimento di Ingegneria Gestionale, Politecnico di Milano, 20156 Milano, Italy (name.surname@polimi.it)
[3] Huawei Technologies Italia S.r.l., 20090 Segrate, Italy (name.surname@huawei.com)
[4] Huawei Technologies France, 92100 Boulogne-Billancourt, France (name.surname@huawei.com)
[5] Samsung Electronics R&D Institute, TW18 4QE Staines, UK (name.surname@samsung.com)
[6] Vodafone Automotive, 00138 Rome, Italy (name.surname@vodafone.com)
[7] Nokia Bell Labs, 70435 Stuttgart, Germany (name.surname@nokia-bell-labs.com)
[8] Intel Deutschland GmbH, 85579 Neubiberg, Germany (name.surname@intel.com)

**Abstract**

*6G technology targets to revolutionize the mobility industry by revamping the role of wireless connections. In this article, we draw out our vision on an intelligent, cooperative, and sustainable mobility environment of the future, discussing how 6G will positively impact mobility services and applications. The scenario in focus is a densely populated area by smart connected entities that are mutually connected over a 6G virtual bus, which enables access to an extensive and always up-to-date set of context-sensitive information. The augmented dataset is functional to let vehicles engage in adaptive and cooperative learning mechanisms, enabling fully automated functionalities with higher communication integrity and reduced risk of accidents while being a sentient and collaborative processing node of the same ecosystem. Smart sensing and communication technologies are discussed herein, and their convergence is devised by the pervasiveness of artificial intelligence in centralized or distributed and federated network architectures.*

**Keywords**: *6G V2X, virtual bus, smart wireless environments, joint communication and sensing, sensor orchestration, AI interface, peer-to-peer computing, federated and coordinated computing.*

## I. VISION ON 6G V2X

Since almost two years, 5G in the frequency-range 1 (FR1, i.e., the sub-6 GHz bands) is being successfully rolled out worldwide. However, millimeter-wave (mmWave) 5G deployed in the frequency-range 2 (FR2: 24-40 GHz) has just started and its Return on Investment (RoI) is not yet clearly understood, as mmWave access has been deployed in a much smaller number of sites, mainly in densely populated areas. Operators and manufacturers of the telecommunication industry are aware that mmWave radio access is a very promising opportunity to provide users with new appealing services, not just for high-performance use cases, but as a technology enabler of smart cities supported by a truly intelligent ecosystem with robots, tactile internet, virtual and augmented reality (VAR), and pervasive intelligent transportation system (ITS) (see Fig. 1). These and more verticals are braced by artificial intelligence (AI) at all levels, accompanied by the ever-growing availability of consistent, secure, and huge exploitable data sets.

The focus of research activities has recently moved towards the next generation of telecommunication system, i.e., 6G. Several key performance indicators (KPIs) are currently debated, but realistically one of the key aspects of 6G will be the shift towards higher spectrum portions such as 61 - 81 GHz (recently agreed), and up to sub-THz (D-band) for a user experience of 10 Gbps and latency < 1 ms, thus being at east 10 times better compared to 5G KPI, as detailed in Fig. 2. 6G will likely guarantee a hyper-connectivity never experienced before and will enrich the 5G ecosystem to enable all those KPI-critical verticals that require extreme connectivity. To this aim, *a beam-type communication is mandatory* to cope with the propagation loss. Narrow beams need to be collimated and beam alignment and tracking is a key issue to be investigated within the scientific community. A promising emerging concept to mitigate this problem and enable 6G propagation is represented by smart electromagnetic (EM) environments: a multitude of passive reconfigurable intelligent surface (RIS), holographic multiple-input multiple-output (MIMO) and active amplify & forward (AF) relays have to be deployed to compensate the lack of line-of-sight (LOS) in any possible condition as detailed in Fig. 1. 6G network will perform coordinated sensing operations on top of simply carrying data for its users according to the joint communication and sensing (JC&S) paradigm.

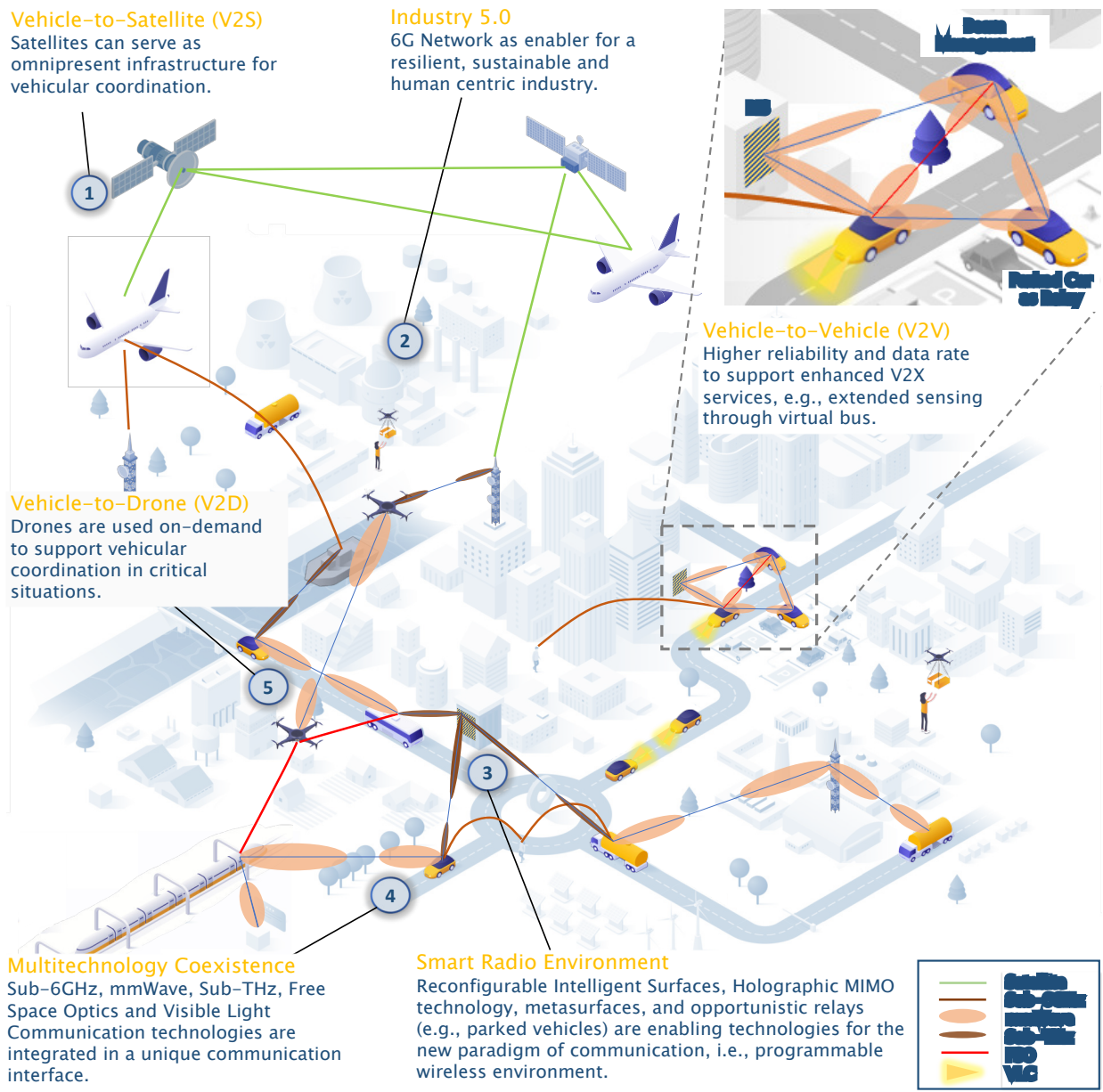

Figure 1: 6G V2X scenario: everything is connected in the wireless 2.0 paradigm

The evolution to 6G and the penetration of connected and autonomous vehicles (CAVs) is an extraordinary timely conjunction of events. Over the next 10 years, the automotive industry will face a revolutionary change where the value chains will shift to a multi-actor ecosystem and the original equipment manufacturer (OEM) will not be the sole key player [1]. The electrification is posing an epochal shift in the value chain as well as the cooperation between vehicles and intelligent mobility is an evolving powerful concept with an enormous amount of on-board electronics that will impact the vehicles' design. This is in conjunction with the fast and massive 6G vehicle-to-everything (V2X) connectivity among vehicles either over infrastructure network (vehicle-to-infrastructure V2I) and among themselves (vehicle-to-vehicle, V2V).

There are many use cases, and related business models, that will be leading the automotive market such as: real-time situational awareness, high-definition 3D maps, cooperative perception, software over-the-air (OTA) updates of the AI on vehicles, tele-operated driving, high-precision positioning, pilotless driving, preventive maintenance, and high-density platooning. In 10+ years customers will change the concept of mobility buying either smart-mobility services or robotaxi rides, rather than real cars as has happened so far now. Self-delivery by autonomous trucks integrated with last-mile transport using unmanned aerial vehicles (UAV) will be another important paradigm for goods delivery ("last mile distribution models").

The rest of the paper cover all these topics, and it is organized as follows: all the above-mentioned visionary concepts with emphasis on sensors orchestration are dealt with in Section 2, 6G V2X services are treated in Section 3, 5G vs 6G capabilities are compared in Section 4, an analysis of CAV enabling technologies is provided in Section 5, and advanced sensing and localization in Section 6.

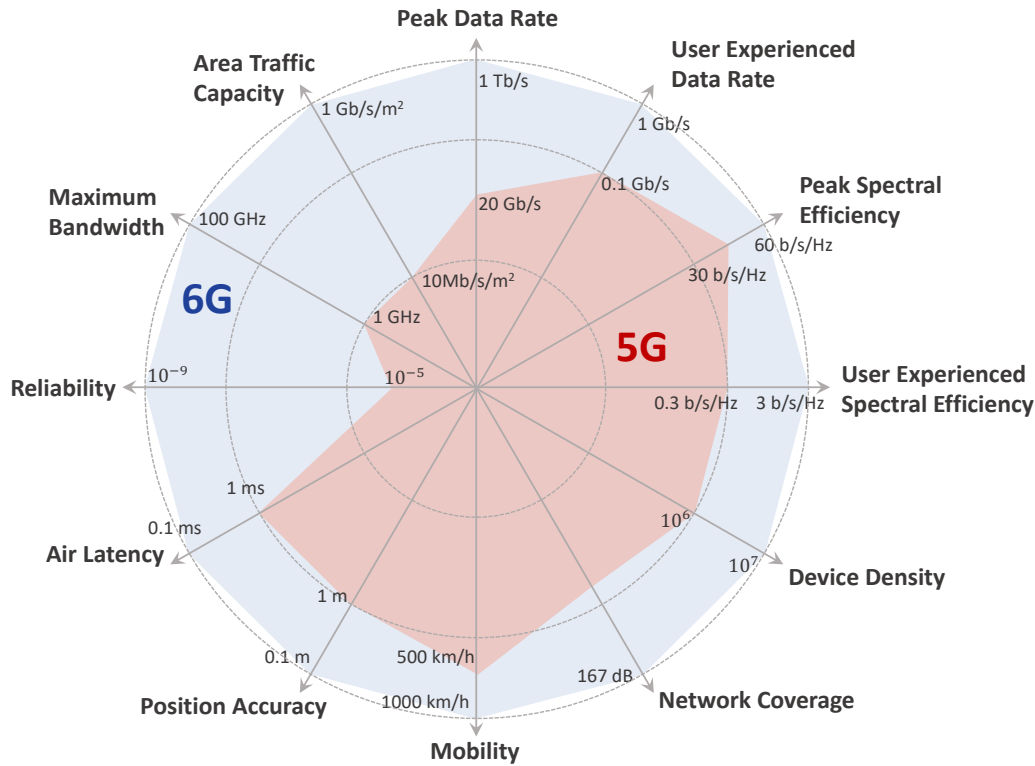

Figure 2: Technical requirements and KPIs of 6G system [2].

## II. SENSOR ORCHESTRATION

6G V2X communications will truly enable the cooperation among fully autonomous driving vehicles (Level of Autonomy – LoA – 4/5). AI assisted cooperative perception of the dynamic environment is crucial for vehicle maneuvering and inter-vehicle sensor fusion for the data coming from other vehicles, pedestrians, or roadside units (RSUs) as well as the growing ecosystem of *smart* infrastructures. This is a major evolutionary breakthrough empowered by massive, ultra-reliable 6G connectivity.

In the mobility ecosystem, a key component is the distributed sensor processing and orchestration. Acquisition of data from global navigation satellite system (GNSS), cameras, lidars, and MIMO radars distributed over multiple road-entities needs to be orchestrated over a fast 6G V2X bus using the principle of parsimony. This connectivity over a distributed smart data-bus connecting all sensors from each vehicle enables each road entity to access the other entity sensors' data that is strictly necessary to augment its environment perception (see Fig. 3). This exchange of *raw sensor data* will enact a new model of cooperation between the automotive and the ICT (information and communications technology) industries in the evolution towards cooperative CAVs as well as *smart*, adaptable, and eco-friendly transportation, and mobility environments.

Sensors' orchestration has a strong legal argument that will impact the insurance systems: in case of any accident, the liability is limited to one (or few) legal entity that possibly actively participated and caused it. This is far better than signaling occupancy grids or any pre-manipulated data that would bring to unpleasantly share the liability also to cooperating entities. 6G-enabled sensing orchestration across multiple fixed/moving platforms will provide an unprecedented Multiview sensing capability for a privacy-preserving AI-fusion on each vehicle moving in urban and highway scenarios. 6G communication waveforms with multi-GHz bandwidth are part of JC&S orchestration of multi-static radar sensing. Orchestrated radar-imaging is organized in term of cooperative MIMO radar, and in-vehicular systems the radars' movement is used in high-resolution synthetic aperture radar (SAR) system with high azimuth resolution < 1 deg [3].

Sensors' orchestration is an enlightening example of the necessary partnership between ICT and automotive industry to pragmatically evaluate and design the prioritization of 6G communication, computation, privacy, and storage for the largest degree of autonomy of vehicles. This industrial cooperation will likely break down the barriers of mutual prejudice and yield to parallel standardization activity over the next decade.

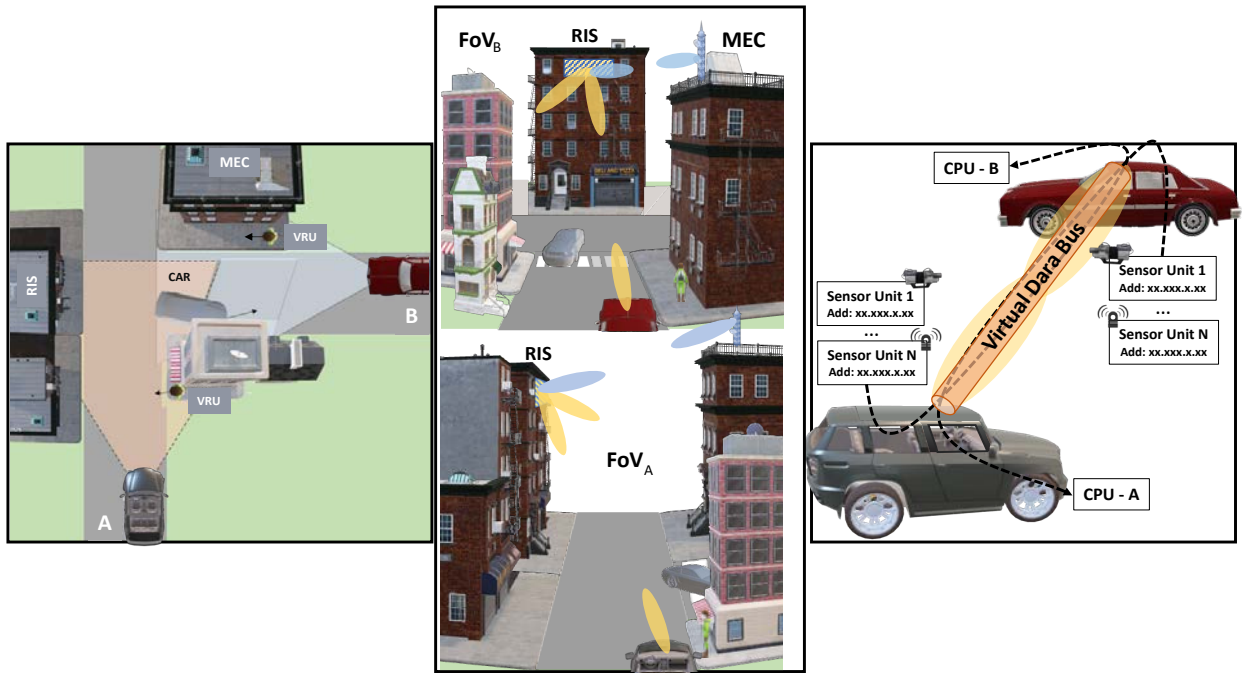

Figure 3: Vehicles A and B using their sensors would not have complete visibility of the scenario, being their Field of View (FoV) limited, as can be observed in the right and central frames. 6G V2X virtual bus would enable vehicles to extend their sensing capabilities by receiving key information from nearby road users, i.e., right frame.

### III.  FULL AUTOMATION V2X SERVICES

6G connectivity between vehicles, infrastructure and road users is the basis for building an enhanced cooperative ITS (C-ITS) platform that provides advanced services to CAVs.

First attempts of C-ITS services date back to the early 2000s, with vehicular *ad-hoc* networks and standardization activities on WLAN-based dedicated short-range communications (DSRC). Technology solutions based on such standard (e.g., ETSI ITS G5 and IEEE WAVE) are today being used to deploy smart road infrastructures providing basic cooperative services for road safety (e.g., collision warning, lane change support, road hazard warnings) and traffic efficiency (e.g., speed management or in-vehicle signage). These early services consist in broadcasting awareness/warning messages with a strong focus on safety. On the other hand, C-V2X (standardized in 2014) and particularly 5G NR Release 16 (in 2020) are expected to support enhanced services such as vehicle platooning, advanced driving, extended sensing, and remote driving. This is thanks to an ultra-reliable low-latency communication (URLLC) which is crucial to enable cooperative perception and maneuvering functionalities [4].

5G NR is unlikely to be able to support fully automated operations (those referred as LoA 5) in complex scenarios within high-density traffic conditions. In full-automation use-cases, in fact, densely CAVs need to continuously share large volume of sensor data for cooperative perception to guarantee safety in all situations. CAVs have to interact in real-time to negotiate the maneuvering and mutually synchronize to achieve optimal driving patterns, with maximized efficiency and minimized environmental impact. A multi-GHz bandwidth zero-latency connectivity is clearly required to let the AV achieve demanding goals in all situations together with a cooperative, peer-to-peer computing and storage meta-model.

Besides the widely acknowledged benefits in safety and efficiency, the convergence of 5G/6G communication and driving automation is going to revolutionize the usage that we make of vehicles. Currently we are living the early phase of the self-driving revolution, where fully AVs are developed by OEM and employed within controlled/private environments (e.g., industrial fleets for logistics, farming, airports, etc.). Over the next decade, the paradigm shift will face the consumers to use CAVs, thus spending traveling time on working, relaxing, or accessing entertainment. Users will largely benefit from intelligent and autonomous transportation, and logistics. The car will evolve from a pure mean of transport to a service platform, an extension of the user's home/office and entertainment environment. The 6G vehicular cloud is the application landscape, with a variety of mobility services hitting the road, such as infotainment, autonomous parking, reporting of traffic conditions and accidents. 6G V2X will enable the sensors' orchestration and distributed AI assisted management of the traffic process. Intra/inter-sensing is orchestrated on the surroundings and vehicles coordinate their motion to maintain traffic smooth and safe. By distributed intelligence, they will move seamlessly through intersections, find the best place to park and the way to deliver passengers and/or goods to their destination with maximum safety/comfort, shorter delivery times, and minimum impact on the environment.

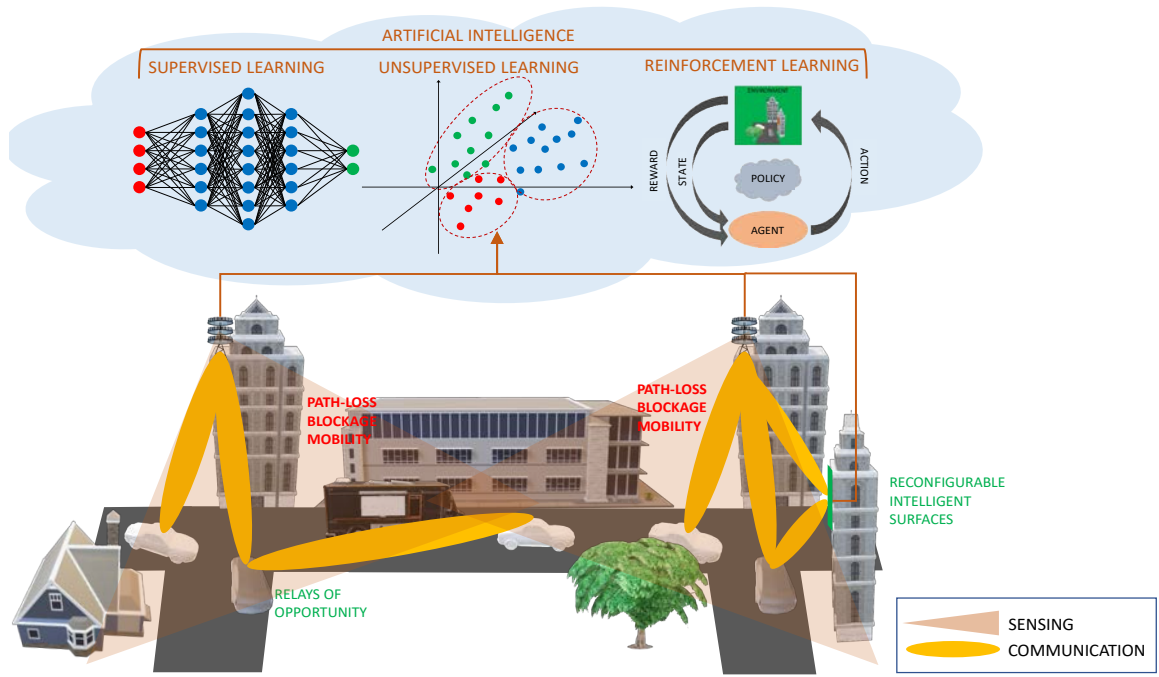

Figure 4: Fully AI-based communication and sensing in the Wireless 2.0 6G V2X scenario: by leveraging AI techniques (supervised, unsupervised and reinforcement learning) and new technologies (relays of opportunity, RIS, IRS, Ho-MIMO) the wireless environment turns into an optimization variable, which, jointly with the transmitters and receivers, can be controlled and programmed.

To reach such a challenging goal, 6G V2X is the technology enabler for vehicles to self-organize into distributed computing and storage platforms that provide the above-mentioned services by collecting, processing, storing, and delivering information in real-time and on a per-needed-basis. Critical time constraints along with local data relevance call for a decentralized processing at the vehicles and cooperation by 6G V2X URLLC.

## IV. 6G vs 5G FOR CAV

The stringent requirements to enable the full potentials of CAVs are based on the innovation behind 6G V2X (Fig.2), which are largely enhanced compared to 5G [5].

The 6G technology will enable object sensing and positioning for high-fidelity 3D holographic reconstruction of the environment from multiple viewpoints orchestration. This requires an extremely accurate synchronization of the events, end-to-end latency of 0.1 ms with jitter targeted below 1 s. The extreme reliability (Fig.2) makes the link connection stable like a *virtual cable* across road entities. 6G will aggregate the interoperation of multiple networks, thus going well beyond today's terrestrial networks. Fig. 1 shows that Low Earth Orbit (LEO) satellites and moving platforms, such as UAV, airplanes, ships, and high-speed trains will be fully integrated into the same ecosystem.

All the above distinctive characteristics are enabled by spectrum above 20 GHz, which has to be scheduled and shared among densely deployed road entities. The huge amount of data signals from communication and orchestrated sensing is efficiently integrated by the AI that is either on-board or distributed. Edge-cloud systems are continuously involved to exchange this learning across multiple platforms to reach a consensus on AI rules (e.g., maneuvering, sensing orchestration, fusion). AI will disruptively impact the design of 6G networks leveraging on the, by then, pervasivity of the 5G FR1 network. The dynamic network settings will be optimized by using AI for the software-based control. The AI will not only impact the control and the stability of the network, but also the design of physical layer functionalities to enhance spectrum utilization or beam management, the reconfiguration of the wireless propagation environment through RIS, metasurfaces, and dynamic relays [6], as delineated in Fig. 4. A dense deployment of RISs and their smart coordination allows to reconfigure the signal propagation between transmitters and receivers to tackle the wireless channel fading impairment and interference issues.

Current standardization activities for the next generation wireless connectivity at mmWave in vehicular networks are based on orthogonal frequency division multiplexing (OFDM) as by 3GPP NR V2X [7]. The race of 3GPP and IEEE802.11 for V2X services has been recently (Nov. 2020) defined by the Federal Communications Commission (FCC) that changed the licensing plan of lower 45 MHz at 5.850-5.895 GHz for unlicensed commercial uses such as WiFi and left the upper 30 MHz for ITS. FCC has endorsed the C-V2X as the standard for automotive uses. The support of V2X by 3GPP is provided over the PC5 interface by sidelink (SL) transmissions. The main characteristics are the use of mini-slot scheduling, which allows to schedule transmission of a slot for latency-critical services. Improved localization of vehicles is made possible using large antenna arrays, and the exchange of safety-based (V2X, ETSI-compliant) messages. Another aspect is the definition of the resource allocation modes 1, where the base station schedules the resources, and 2, which lets the user equipment (UE) autonomously select the SL transmission resources. In summary, we expect that 6G will be focused on *(i)* upgrade of SL evaluation methodology; *(ii)* low-power and low-latency resource allocation enhancement, particularly for mode 2; *(iii)* SL discontinuous reception options for unicast, groupcast, and broadcast; and *(iv)* addition of new SL frequency bands, including FR2-specific enhancements

With respect to the physical layer, the main aspects in the transition from 5G to 6G will be on the numerology and on all related to sensor orchestration that can be distributed with tight latencies. For V2X systems, the fast association/disassociation with RSUs may require a distributed antenna deployment with 10-100 GHz bandwidth and strict air-interface jitter (<1 ns).

## *V. ENABLING V2X TECHNOLOGIES FOR SMART CAV ENVIRONMENT*

6G V2X will leverage a jumble of different technologies, mutually harmonized to guarantee the requirements of data rate, latency, and degree of perception of the surrounding environment. The development of new communication and sensing technologies will make a transition from channel-based design to programmable propagation environments, thus realizing a smart radio environment (SRE), also named Wireless 2.0. The concept is portrayed in Fig. 4. In the following, we review the expected technological advancements to enable 6G, outlining their main features and applications.

### A. High-frequency Communications

Communication systems operating in the mmWave (30 -100 GHz), sub-THz (100 - 300 GHz), THz (300 GHz - 10 THz) and optical (> 10 THz) bands are envisioned to be pillars of 6G V2X networks. While mmWave is on the verge of being technologically consolidated, sub-THz and THz systems will experience a rapid growth in the next years, mostly motivated by the recent advancements in sub-THz and THz devices and packaging, filling the *THz gap* between traditional radiofrequency and optical systems. In this sense, 6G V2X could be enabled by large sub-THz and THz unlicensed bandwidths (e.g., 7 GHz at 120 GHz carrier frequency), possibly sliced for multiple access. The mm or sub-mm wavelength supports the propagation using "near-pencil beams", either by packing hundreds (mmWave) to thousands (sub-THz) of small-footprint antennas in compact ultra-massive MIMO (UM-MIMO) arrays, or by leveraging dielectric lenses to collimate the emitted beam with less technological effort compared to conventional beamforming structures [6], [8]. Piggybacked to communication purposes, high-frequency sensing allows to probe the surrounding environment with unprecedented cm- or sub-cm accuracy, for high-precision simultaneous localization and mapping (SLAM), thus enabling 3D environment imaging. The rush to THz systems has been enabled by technological advancements in complementary metal-oxide-semiconductor (CMOS) and III-V technologies (InP, GaAs, etc.) used for sub-THz bands, while hybrid electronic-photonic systems are emerging as promising solutions for higher frequencies.

Several results have proved the feasibility of sub-THz multi-Gbps links (> 20 Gbps) at variable ranges, mostly limited to few meters (< 10 m) unless very high-gain sub-THz arrays (> 40 dBi) are employed. Although these achievements represent a strong basis for short-range (< 50 m) sub-THz/THz V2X links, numerous challenges are expected to be solved before mass-market development. From one side, technological challenge will be the optimization and integration of sub-THz and THz devices (e.g., oscillators, modulators, phase shifters, feed networks) and their packaging. On the other side, sub-THz and THz systems exacerbate the propagation issues arose at mmWaves, such as the path-loss, which is typically two orders of magnitude higher. Sub-THz/THz wireless channel is mostly composed by a LOS path, and adverse weather (e.g., rain, fog) affects proportionally to the carrier frequency and relatively long distances (>100 m). As for mmWaves, the main issues for sub-THz and THz systems are *(i)* the random blockages of the LOS path by moving obstacles and *(ii)* the beam management in high-mobility V2X systems. The first encourages the usage of novel technologies augmenting the resilience of the V2X network RIS and relays to avoid non-line-of-sight (NLOS) conditions, while the second leverages on dynamic beam alignment and channel estimation procedures suited for UM-MIMO systems [9].

Incoherent optical wireless communication (OWC) is considered a mature technological option for mid-range (< 200 m) 6G communications [10]. Free-space optics (FSO) is a high-directivity (mrad beamwidth) laser-based technology (typically in the near infrared 850-1550 nm) for multi-Gbps at comparably low cost. While the weather-related issues can be alleviated by short range applications, the implementation in 6G V2X requires a dynamic closed-loop beam control by arrays of micro-electro-mechanical system (MEMS)-based micro-mirrors, possibly equipped with lens-based focusing/de-focusing systems to steer and control the laser beams, but research of alternative solutions is on the way. In dense FSO networks, the eye-safe operation must be guaranteed, and the interference to/from lidar systems must be properly tackled adopting the optical JC&S. Visible light communications (VLC), instead, reuse existing LED lamps as transmitters and either photodetectors or cameras as receivers, allowing up to 100+ Mbps links over tens of meters distance on the visible portion of the EM spectrum (380-780 nm). VLC is envisioned to serve in 6G V2X networks exploiting vehicles' front and rear lights in V2V links as well as traffic lights and light poles for V2I communications. Overall, optical technologies will synergize with 6G radios for augmented reliability.

### B. Resilient Network Technologies (RIS, IRS, holographic MIMO, smart relays)

Beam-management and link-blockage are challenging issues for 6G V2X, to address which recent advancements on RIS, programmable metasurfaces, passive holographic MIMO (Ho-MIMO) as well as on dynamic relaying [11] come into help. Active relays, possibly full duplex, provide a cost-effective solution to the V2X scenario by enhancing the robustness and reliability of the vehicular network. Vehicles equipped with a wireless interface can also serve as relays of opportunity when they are not engaged in their own transmissions, e.g., parked cars while recharging. Ho-MIMO [12] is a novel low-cost and low-power dynamic beamforming technique. Each antenna is connected to a single varactor, which defines the impedance of the antenna. The direction of radiation is controlled by changing the impedance of each antenna, and the pattern of impedances is the holographic configuration.

Differently, an RIS is a two-dimensional material structure that is reconfigurable in terms of its EM wave response, such that: *i)* reflect/refract an impinging radio wave towards any direction; *ii)* modify the polarization of the incident radio wave; *iii)* focus/collimate the incident waves from/towards a predefined location. The RIS increases channel rank with "artificial" controllable paths, coverage, and interference suppression. 6G V2X blockage of propagation is largely reduced by RIS with an increase of the data rate, thus achieving higher reliability and Quality of Service (QoS). A lower-cost and -complexity alternative to RISs is represented by intelligent reflecting surfaces (IRSs) or metasurfaces. As opposed to RISs, IRSs are not reconfigurable in real-time, which decreases the flexibility of the EM environment and reduces the complexity and deployment cost. A careful design could explore the full potential of IRSs. For example, future CAVs can be equipped with IRSs on their sides, which could enrich the scattering in dense road traffic and increase the channel rank with remarkable benefits on blockage.

### C. Smart V2X Protocol Architectures

6G V2X evolution needs to support mass deployment of advanced safety automated driving use cases and high-definition sensors sharing. These complex mission critical interactions will require flexible computing and communication frameworks supporting new emerging technological trends.

Proactive and predictive system approach: automotive applications will be dependent on 6G extreme performance. These applications and services are part of Business-to-Business (B2B) or Business-to-Business-to-Customer (B2B2C) contracts, any unpredicted service quality issues might result in penalty clauses or liability for service providers. Predictive QoS enables to deliver information about QoS changes that are likely to occur to make such B2B or B2B2C applications viable. Industry has already started some preliminary work on this topic.

Resources' virtualization: in Beyond 5G and 6G systems, specific network functions and resources are virtualized. Network architecture will evolve towards Multi-Access Edge Computing (MEC) distributing cloud applications closer to the user. Future in-vehicle applications such as truly immersive cross-reality, or mobile holograms require extensive computing to deliver real-time immersive user experiences. These applications may leverage on split computing to make use of reachable computing resources over the network and MEC servers.

Functional safety requirements in V2X: to address potential hazards emerging from connectivity failure, or its reduced performance, the V2X reliability requirements will become more complex. According to the vehicle functional safety standard ISO26262, a risk and hazard analysis determines Automotive Safety Integrity Level (grade A-D) by weighting potential to threaten lives (highest grade D). There is a need to identify new intelligent service reliability in 6G V2X architecture mechanisms to fulfil increased functional safety requirements in radio resource management and core network such as multi-connectivity, link redundancies, QoS handling, offloading, MEC, and network slicing.

## VI. SENSING AND LOCALIZATION TRENDS

6G V2X connectivity enables the orchestration of sensing by sharing the measurements over a common virtual bus (see Sec. II). 6G is the bridge for information sharing, outsourcing locally generated information and receiving from external proximity devices or the intelligent network itself. The augmented dataset allows each autonomous entity to take proactive decisions, with a higher integrity and reduced risks, and engage in adaptive and cooperative learning mechanisms.

### A. Joint Communication and Sensing Networks

JC&S is the emerging trend in 6G research [13], whose aim is to deliver a common platform to perform sensing and communication. The available spectrum for V2X will be then leveraged to perform environment scanning operations typical of radars, on top of existing running communications. This would allow 6G to occupy an even more central role than 5G, given its new sensing task, typically demanded for ad-hoc solutions.

Recent works, like [2], start to address the trade-off between assigning spectrum to communications and sensing operations. While defining the best physical co-existence schemes between these two operating modes is one of the challenges of 6G, it is already clear from now that the network could reach congestion in highly trafficked areas, where vehicles orchestrate sensing and legacy communication services and different best effort applications, like video streaming.

State of the art QoS and Quality of Experience (QoE) definitions for communication services must be enhanced to include their correspondent for sensing operations. From rate, reliability and latency guarantees, the network should find ways to guarantee enough sensing precision, reliability, latency, jitter, and information update rates, just to mention some of the possible KPIs of interest Fig.2. Accordingly, all existing layer 2 and above mechanisms for the next generation of JC&S wireless networks must be re-designed. Among them, the proper design of admission control, radio resource allocation, and congestion resolution schemes will be pivotal for a successful deployment of 6G JC&S networks.

### B. Hybrid Positioning

In urban environments, the high density of objects (buildings, trees, billboards, signs, parked vehicles, etc.) obstructs both sensing and communication over the LOS. However, the envisaged massive deployment of connected sensors and distributed intelligence will significantly create an overabundance of information that can be accurately managed and integrated to relax the LOS requirement. An emerging concept for urban mobility highly recommends the hybridization of positioning technologies. In a first phase, satellite-based positioning (i.e., GNSS), which greatly suffers in urban areas, is deemed to be complemented with cellular-based localization through temporal or spatial measurements. Not surprisingly, 3GPP 5G NR Release 16 has introduced the positioning reference signals for localization purposes. In a second phase, the pervasive deployment of an unprecedented number of 6G devices (envisioned density of $10^7$ 6G connections per $km^2$, see Fig. 2) in cities will enable precise 6D positioning (i.e., 3D position and 3D orientation) without needing the historically conceived pivotal satellite-based solution. 6G-based localization will be the first technology with coverage, robustness, redundancy and, most importantly, precision features to enable the diffusion of vehicular autonomy at large scale, despite the extremely severe requirements on positioning accuracy (< 10 cm) for CAVs on LoA 5.

### C. Processing Architectures

As pointed out in previous sections, a pervasive spread of 6G connected entities raises the research and industry attention on cooperative techniques for data processing, with major efforts to design efficient and scalable architecture that optimize the resources and investments [14]. In this regard, centralized and distributed architectures are feasible. Also, hybrid versions are possible, with the centralization of only a subset of tasks. In the centralized case, the role of the 6G network is pivotal, as it must allow for a low latency data aggregation, fast processing, and flooding. In case of fully distributed networks, the connected entities bypass the network for a faster and local processing by direct communications (such as V2V), where URLLC 6G guarantees near real time awareness and tactile feedbacks. In hybrid architectures, the tradeoff between delegating dedicated processing tasks to the 6G edge and locally retaining all the computational burden leads to an increased attention to edge computing and distributed ledger technologies, which are foreseen to become more and more popular in the next future. The choice of network architecture also leads to the design of different algorithms, with specific requirements in terms of sensor orchestration and processing, computational power, communication, and energy consumption for electrical mobility.

In this context, currently used AI techniques, such as supervised, unsupervised and reinforcement learning, will play a pivotal role, but the recently emerged federated learning concept will begin to take hold. Federated machine learning [15] is emerging as key technology for cooperative CAVs that perform several automated functions relying on bigdata. The integration of federated learning techniques with vehicles acting as distributed learners is expected to enable faster, more accurate and flexible training as well as novel decision-making opportunities. Federated learning enables decentralized cooperative fusion and training, without requiring raw data sharing, with huge benefits in latency and in preserving the users' privacy.

## VII. CONCLUSIONS

In this paper, we draw the envisioned connected mobility scenario where 6G communication technology will operate. We depict the 6G V2X vision as a whole, discussing the major trends on mobility applications and services. Major attention is given to the concept of sensor orchestration by 6G virtual bus, where all road entities become part of a wider connected intelligence. The discussion also embraces the enabling technologies that allow a transformation of the environment into an active and programmable entity through pervasive AI. The radical transformation of the intelligent 6G network introduces new research areas and trends, where communication and sensing combine over centralized or distributed architectures. The discussed topics currently represent the hottest trends in research. They also open new possibilities for synergies and collaborations between automotive and ICT industries for the design of a truly smart, proactive, and connected V2X mobility ecosystem.